\documentclass[12pt,epsfig]{article}

\usepackage{graphicx,color,here}
\usepackage{here}

\newcommand{\be}{\begin{equation}}
\newcommand{\ee}{\end{equation}}
\newcommand{\bea}{\begin{eqnarray}}
\newcommand{\eea}{\end{eqnarray}}
\newcommand{\PR}[1]{{\it Phys.\ Rev.}\ {\bf #1}}
\newcommand{\PRL}[1]{{\it Phys.\ Rev.\ Lett.}\ {\bf #1}}
\newcommand{\PL}[1]{{\it Phys.\ Lett.}\ {\bf #1}}

\newcommand{\bfk}{\mbox{\boldmath $k$}}

\newcommand{\bfP}{\mbox{\boldmath $P$}}
\newcommand{\bfS}{{\mbox{\boldmath $S$}}_{_T}}

\newcommand{\pup}{p^\uparrow}
\newcommand{\pdown}{p^\downarrow}

\newcommand{\bfp}{\mbox{\boldmath $p$}}

\def\bkt{\bf k_\perp}
\def\bpt{\bf p_\perp}
\def\kt{k_\perp}
\def\pt{p_\perp}

\begin{document}

\title{\bf Cahn and Sivers effects in the target fragmentation region of SIDIS}

\author{Aram~Kotzinian \\[1.cm] \it
 Dipartimento di Fisica Generale, Universit\`a di Torino \\
         and \it INFN, Sezione di Torino, Via P. Giuria 1, I-10125 Torino, Italy\\
and \it Yerevan Physics Institute, 375036 Yerevan, Armenia \\
and \it JINR, 141980 Dubna, Russia\\
\it email: aram.kotzinian@cern.ch}

\maketitle

\begin{abstract}

{\tt LEPTO} event generator is modified to describe the azimuthal
modulations arising from Cahn and Sivers effects. The comparisons
with some existing data in the current fragmentation region of SIDIS
are presented. The predictions for Cahn and Sivers asymmetries in
the target fragmentation region are presented for SIDIS of 12 GeV
electrons off proton target.
\end{abstract}

\section{Introduction\label{sec:intro}}

\begin{figure}[h!]
\begin{center}
\scalebox{0.4}{\input{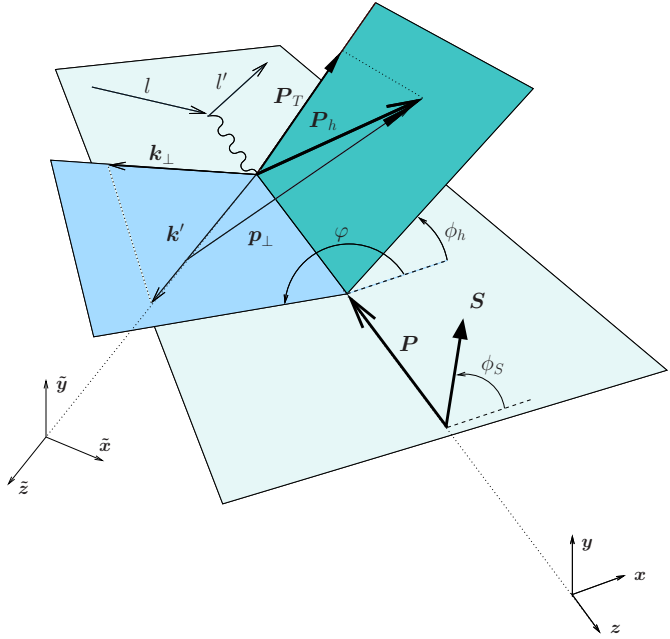}} \caption{\small Three
dimensional kinematics of the SIDIS process.}
\label{fig:planessidis}
\end{center}
\end{figure}

In~\cite{akp} the role of parton intrinsic motion in
semi-inclusive DIS (SIDIS) processes within QCD parton model has
been considered at leading order; intrinsic $\bkt$ is fully taken
into account in quark distribution functions and in the elementary
processes as well as the hadron transverse momentum, $\bpt$, with
respect to fragmenting quark momentum, see
Fig.~\ref{fig:planessidis}\footnote{In the following the notations
of~\cite{akp} are used.}.
%
%
The average values of $\kt$ for quarks inside protons and $\pt$
for final hadrons inside the fragmenting quark jet where fixed by
a comparison with data on Cahn effect~\cite{cahn} -- the
dependence of the unpolarized cross section on the azimuthal angle
between the leptonic and the hadronic planes. The single spin
asymmetry (SSA) $A_{UT}^{\sin(\phi_\pi - \phi_S)}$ recently
observed by HERMES Collaboration~\cite{hermUT} was successfully
described by Sivers mechanism~\cite{siv}. It was also shown that
the Sivers distribution functions resulting from this analysis are
compatible with the preliminary data from COMPASS
collaboration~\cite{compUT}.

The description of SIDIS within standard QCD parton model approach
using the distribution and fragmentation functions is valid only
in the current fragmentation region, CFR ($x_F>0$) and at high
energies. A more general approach allowing to describe SIDIS in
the whole kinematic region available for final hadrons is based on
the LUND string fragmentation model and is incorporated into {\tt
LEPTO} event generator~\cite{lepto}. In the simplest case,
corresponding to LO approximation of parton model, event
generation in {\tt LEPTO} proceeds in several steps:
\begin{enumerate}
\item The active quark inside the nucleon is chosen
according to the quark density function $f_q(x,Q^2)$,
\item The hard scattering kinematics is generated,
\item The transverse momentum of the final quark is simulated with Gaussian $\kt$
and flat $\varphi$ distributions. Note that the transverse
momentum of the final final quark is equal to that of initial
quark for leading order hard subprocesses.
\item The string fragmentation machinery of {\tt JETSET}
program~\cite{JETSET} is applied to form the final hadrons.
\end{enumerate}

Within this approach the SIDIS cross section at LO can be
expressed as
\be
    \label{sidis}
    \frac{d^5\sigma^{\ell p \to \ell h X }}{dx \, dQ^2 \, dx_F \,
    d^2 \bf P_T} = \sum_q \int {d^2 \bfk _\perp}\; f_{q}(x,k_\perp)
    \; \frac{d \hat\sigma ^{\ell q\to \ell q}}{dQ^2} \;
    H_{q/N}^h(x,x_F,\bfk_\perp, \bf P_T),
\ee
where $\frac{d\hat\sigma ^{\ell q\to \ell q}}{dQ^2}$ is the
lepton--quark hard scattering cross section and
$H_{q/N}^h(x,x_F,\bfk_\perp, \bf P_T)$ represents the
hadronization function of the system formed by struck quark with
transverse momentum $\bkt$ and target remnant. In the standard
version of {\tt LEPTO} the quark distribution function and the LO
lepton--quark cross section are independent of $\varphi$, and,
thus the final quarks are uniformly distributed in azimuthal
angle. For final hadrons this implies also a uniform azimuthal
distribution. However, already in unpolarized SIDIS the observed
azimuthal distribution of hadrons is not flat.

In this paper two types of azimuthal modulation at the quark level
and their influence on the produced hadron azimuthal distribution
will be considered:
\begin{itemize}
\item{azimuthal modulation of the hard scattering cross section in
unpolarized SIDIS (Cahn effect)}
\item{azimuthal modulation of the initial quark transverse momentum
in SIDIS of unpolarized leptons off the transversely polarized
nucleon (Sivers effect)}.
\end{itemize}
It is possible to incorporate both effects in the {\tt LEPTO}
event generator and obtain predictions for azimuthal asymmetries
in the whole kinematical region for the final hadrons. The way how
the {\tt LEPTO} code is modified to include Cahn and Sivers
effects are described in Sec.~\ref{sec:cahn} and
Sec.~\ref{sec:sivers}, respectively. In Sec.~\ref{sec:concl} some
discussion and conclusions are presented.

\section{Including Cahn effect in LEPTO\label{sec:cahn}}

The Cahn effect~\cite{cahn} is a kinematical effect arising due to
the presence of nonzero intrinsic transverse momentum of quarks in
the nucleon. The underlying physics is very simple. The
lepton--quark scattering cross section is given by  the QED
expression
\be
    d\hat\sigma^{\ell q\to \ell q} \propto \hat s^2+\hat u^2.
    \label{lqXsec}
\ee
In the general case of non collinear kinematics Mandelstam
variables depend on the quark transverse momentum and its
azimuthal angle and at order ${\cal O}(\kt/Q)$ one has
\bea
    \hat s^2 &=& \frac{Q^4}{y^2} \left( 1 - 4 \frac{\kt}{Q}\,
    \sqrt{1-y} \, \cos\varphi \right), \nonumber \\
    \hat u^2 &=& \frac{Q^4}{y^2} \, (1-y)^2 \left( 1 - 4 \frac{\kt}{Q}
    \, \frac{\cos\varphi}{\sqrt{1-y}} \right). \label{eq:mand}
\eea
Then, the lepton--quark elastic scattering cross section is given by
\be
    d\hat\sigma^{\ell q\to \ell q} \propto
    1-\frac{(2+y)\sqrt{1-y}}{1+(1-y)^2}\frac{\kt}{Q}\cos\varphi.
    \label{lqcahn}
\ee
Eq. (\ref{lqcahn}) shows that the azimuthal angle of the final
quark (and of the string's end associated with the struck quark)
is now modulated with amplitude depending on $y, Q$ and $\kt$.

This effect can be introduced in the {\tt LEPTO} event generator
at the step 3) of the event generation, when the transverse
momentum and azimuthal angle of the scattered quark are generated.
To do this the generation of the quark transverse momentum, $\kt$,
is left unchanged and then the azimuthal angle is generated
according to Eq. (\ref{lqcahn}). This leads to azimuthal
modulation of the string axis (axis $\tilde z$ on
Fig.~\ref{fig:planessidis}). The momentum conservation means that
the transverse momentum of the quark is balanced by that of the
target remnant, which in turn means that the azimuthal angle of
the target remnant $\varphi_{qq}=\varphi+\pi$. Then, one expects
that the azimuthal angle of the hadrons in the target
fragmentation region (TFR), $x_F<0$, will be modulated with
shifted a phase by $\pi$ with respect to that in CFR.
\begin{figure}[h!]
\begin{center}
\vspace {-0.5cm}
 \includegraphics[width=0.65\linewidth, height=0.55\linewidth]{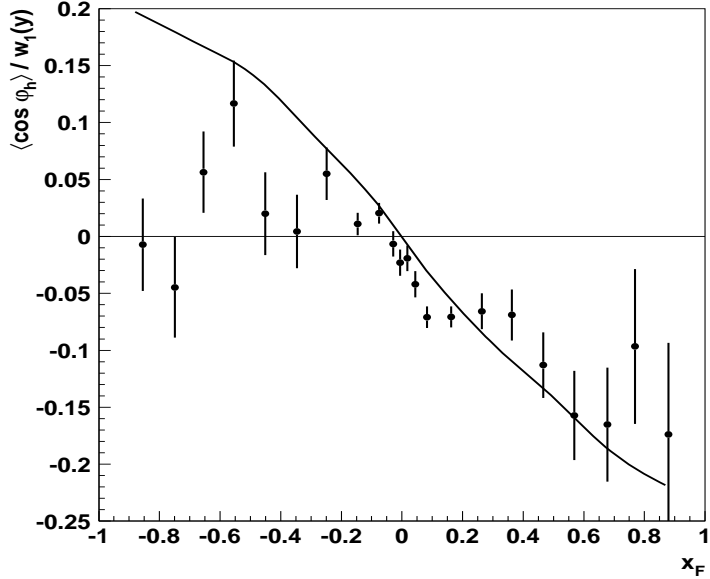}
\vspace {-0.7cm} \caption{\label{fig:cahnemc} {\small The $x_F$
dependence of $\langle \cos \phi_h \rangle / w_1(y)$ for charged
hadrons compared with EMC data.}}
\end{center}
\vspace {-0.5cm}
\end{figure}
\begin{figure}[h!]
\begin{center}
\vspace {-0.5cm}
 \includegraphics[width=0.75\linewidth, height=0.65\linewidth]{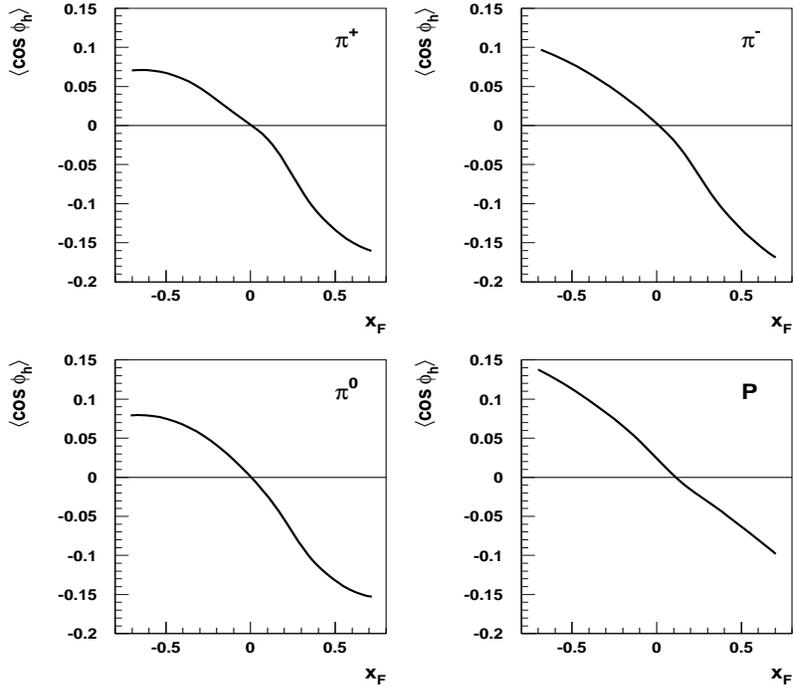}
\vspace {-0.7cm} \caption{\label{fig:cahnjlab1} {\small
Predictions of modified {\tt LEPTO} for $x_F$ dependence of
$\langle \cos \phi_h \rangle$ for different hadrons produced in 12
GeV unpolarized SIDIS process.}}
\end{center}
\vspace {-0.5cm}
\end{figure}

Data on azimuthal dependencies of SIDIS covering a large $x_F$
range have been obtained by the EMC Collaboration~\cite{emc} for a
beam energy of 280 GeV. The $x_F$ dependence of $\langle \cos
\phi_h \rangle /w_1(y),$ where
$w_1(y)=(2-y)\sqrt{1-y}/\left(1+(1-y)^2\right)$, obtained by using
modified {\tt LEPTO} for EMC kinematics are presented in
Fig.~\ref{fig:cahnemc} together with data points from~\cite{emc}.
The simulations has been done with LO setting of {\tt LEPTO}
(LST(8)=0) and with values of the parameters describing intrinsic
$k_T$ (PARL(3)=0.5) and fragmentation $p_T$ (PARL(21)=0.45) as
adopted in~\cite{akp}.

The predictions of modified {\tt LEPTO} for $\langle \cos \phi_h
\rangle$ of different hadron ($\pi^+$, $\pi^-$, $\pi^0$ and $p$)
produced in SIDIS on a proton target at future CEBAF 12 GeV
facility at JLab~\cite{jlab12} are presented in
Fig.~\ref{fig:cahnjlab1}. One can see from Fig.~\ref{fig:cahnemc}
and Fig.~\ref{fig:cahnjlab1} that the predicted mean value of
$\cos \phi_h$ in the CFR is negative $\langle \cos \phi_h
\rangle_{CFR}<0$, while in the TFR is positive $\langle \cos
\phi_h \rangle_{TFR}>0$, as suggested by arguments based on
transverse momentum conservation.

\section{Including Sivers effect in LEPTO\label{sec:sivers}}

The azimuthal modulation of the string transverse momentum in the
previous section was due to Cahn effect -- the dependence of the
non planar lepton-quark scattering cross section on the quark
azimuth. The quark distribution, $f_q(x,k_\perp)$ itself is
independent of quark azimuthal angle.

The situation is different when one considers SIDIS on a
transversely polarized nucleon. Now a correlation between
transverse momentum of quark and target transverse polarization is
possible -- the so called Sivers effect~\cite{siv}. For quite some
time it was believed that this correlation is forbidden because of
T-invariance of the strong interactions. However the spectator
model calculations~\cite{bhs} demonstrated that there exists a
nonzero SSA in SIDIS when the final state interaction between
struck quark and target remnant is taken into account. Then, the
effective description of this SSA is possible within QCD
factorized approach by introducing a new distribution function --
the Sivers function (for further discussion see, for
example,~\cite{metz}).

The unpolarized quark (and gluon) distributions inside a
transversely polarized proton (generically denoted by $\pup$, with
$\pdown$ denoting the opposite polarization state) can be written
as:
\be
    f_ {q/\pup} (x,\bfk_\perp) = f_ {q/p} (x,\kt) + \frac{1}{2} \,
    \Delta^N f_ {q/\pup}(x,\kt)  \; {\bfS} \cdot (\hat {\bfP} \times
    \hat{\bfk}_\perp)\; , \label{poldf}
\ee
where $\bfP$ and $\bfS$ are respectively the proton momentum and
transverse polarization vector, and $\bfk_\perp$ is the parton
transverse momentum; transverse refers to the proton direction.
Eq. (\ref{poldf}) implies
\bea
    \nonumber &&f_ {q/\pup} (x,\bfk_\perp) + f_ {q/\pdown}
    (x,\bfk_\perp) =
    2 f_ {q/p} (x,\kt)\;, \\
    &&f_ {q/\pup} (x,\bfk_\perp) - f_ {q/\pdown} (x,\bfk_\perp) =
    \Delta^N f_ {q/\pup}(x,\kt)\;{\bfS} \cdot (\hat{\bfP}  \times \hat
    {\bfk}_\perp)\;, \label{sivf}
\eea
where $f_ {q/p} (x,\kt)$ is the unpolarized parton density and
$\Delta^N f_ {q/\pup}(x,\kt)$ is referred to as the Sivers function.
Notice that, as requested by parity invariance, the scalar quantity
$\bfS \cdot (\hat{\bfP} \times \hat {\bfk}_\perp)$ singles out the
polarization component perpendicular to the $\bfP-\bfk_\perp$ plane.
For a proton moving along $-z$ and a generic transverse polarization
vector $\bfS = |\bfS|\,(\cos\phi_S, \sin\phi_S, 0)$ (see Fig.
\ref{fig:planessidis}) one has:
\be
    \bfS \cdot (\hat{\bfP}  \times \hat{\bfk}_\perp) = |\bfS| \,
    \sin(\varphi-\phi_S) \equiv |\bfS| \, \sin\phi_{Siv} \,,
\ee
where $(\varphi-\phi_S) = \phi_{Siv}$ is the Sivers
angle.

In~\cite{akp} the Sivers function for each light quark flavor $q=u,d$
are parameterized in the following factorized form:
\be
    \Delta^N f_ {q/\pup}(x,\kt) = 2 \, {\cal N}_q(x) \, h(\kt) \,
    f_ {q/p} (x,\kt)\; , \label{sivfac}
\ee
where
\bea
    &&{\cal N}_q(x) =  N_q \, x^{a_q}(1-x)^{b_q} \,
    \frac{(a_q+b_q)^{(a_q+b_q)}}{a_q^{a_q} b_q^{b_q}}\; ,
    \label{siversx} \\
    &&h(\kt) = \sqrt{2e} \, \frac{\kt} {M} \, e^{-\kt^2/M^{2}}\;  ,
    \label{siverskt}
\eea
where $N_q$, $a_q$, $b_q$ and $M$ (GeV/$c$) are parameters. Then
Eq.~(\ref{poldf})can be rewritten as
\be
    f_ {q/\pup} (x,\bfk_\perp) = f_ {q/p} [x,\kt)(1+|\bfS|{\cal N}_q(x)h(\kt) \,
    \sin\phi_{Siv}].
    \label{sivmod}
\ee

Again, the Sivers effect is incorporated into {\tt LEPTO} at the
stage 3) of the event generation in the same way as for the Cahn
effect but now the azimuthal angle is generated according to
Eq.~(\ref{sivmod}). For simulations the following set of
parameters compatible with those obtained in~\cite{akp} have been
used: $N_u=N_{\bar u}=0.5$, $N_d=N_{\bar d}=-0.2$, $a_q=0.3$,
$b_q=2$ and $M^2=0.36$ (GeV/c)$^2$.

\begin{figure}[h!]
\begin{center}
\vspace {-0.5cm}
 \includegraphics[width=0.75\linewidth, height=0.65\linewidth]{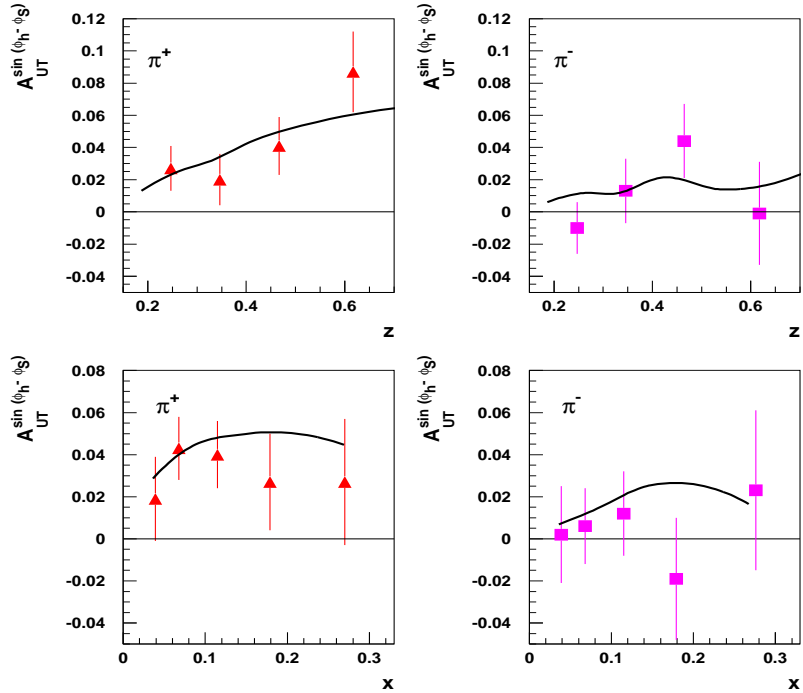}
\vspace {-0.7cm} \caption{\label{fig:sivherm} {\small HERMES data on
$A_{UT}^{\sin(\phi_\pi-\phi_S)}$ \cite{hermUT} for scattering off a
transversely polarized proton target. The curves are the results of
simulations obtained with modified {\tt LEPTO}}.}
\end{center}
\vspace {-0.5cm}
\end{figure}

In Fig.~\ref{fig:sivherm} the results of simulation for HERMES
experimental conditions are compared with observed Sivers
asymmetries~\cite{hermUT}.

\begin{figure}[h!]
\begin{center}
\vspace {-0.5cm}
 \includegraphics[width=0.75\linewidth, height=0.65\linewidth]{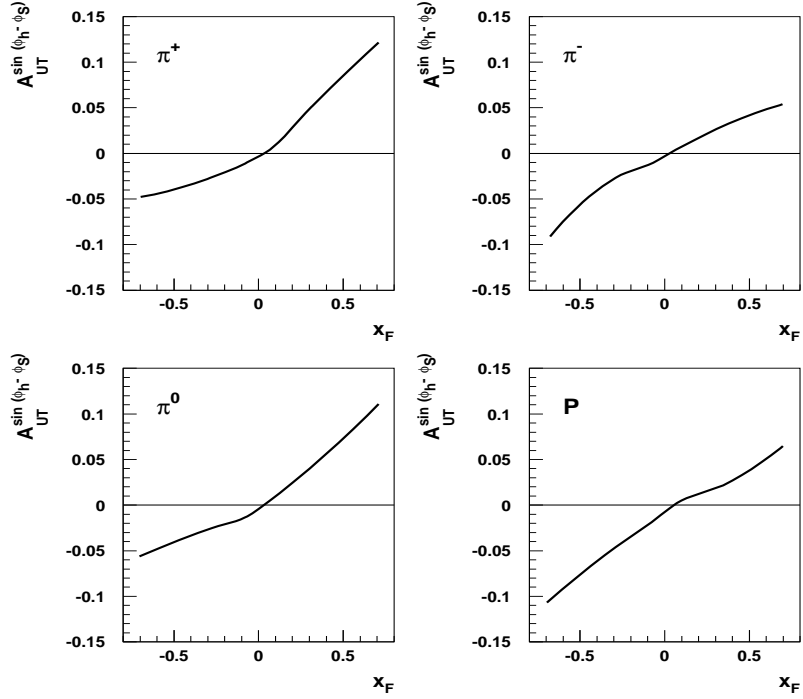}
\vspace {-0.7cm} \caption{\label{fig:sivxf} {\small Predicted
dependence
    of $A_{UT}^{\sin(\varphi_h-\varphi_S)}$ on $x_F$ for different
    hadrons produced in SIDIS of 12 GeV electrons off a transversely polarized
    proton target.}}
\end{center}
\vspace {-0.5cm}
\end{figure}

\begin{figure}[h!]
\begin{center}
\vspace {-0.5cm}
 \includegraphics[width=0.75\linewidth, height=0.65\linewidth]{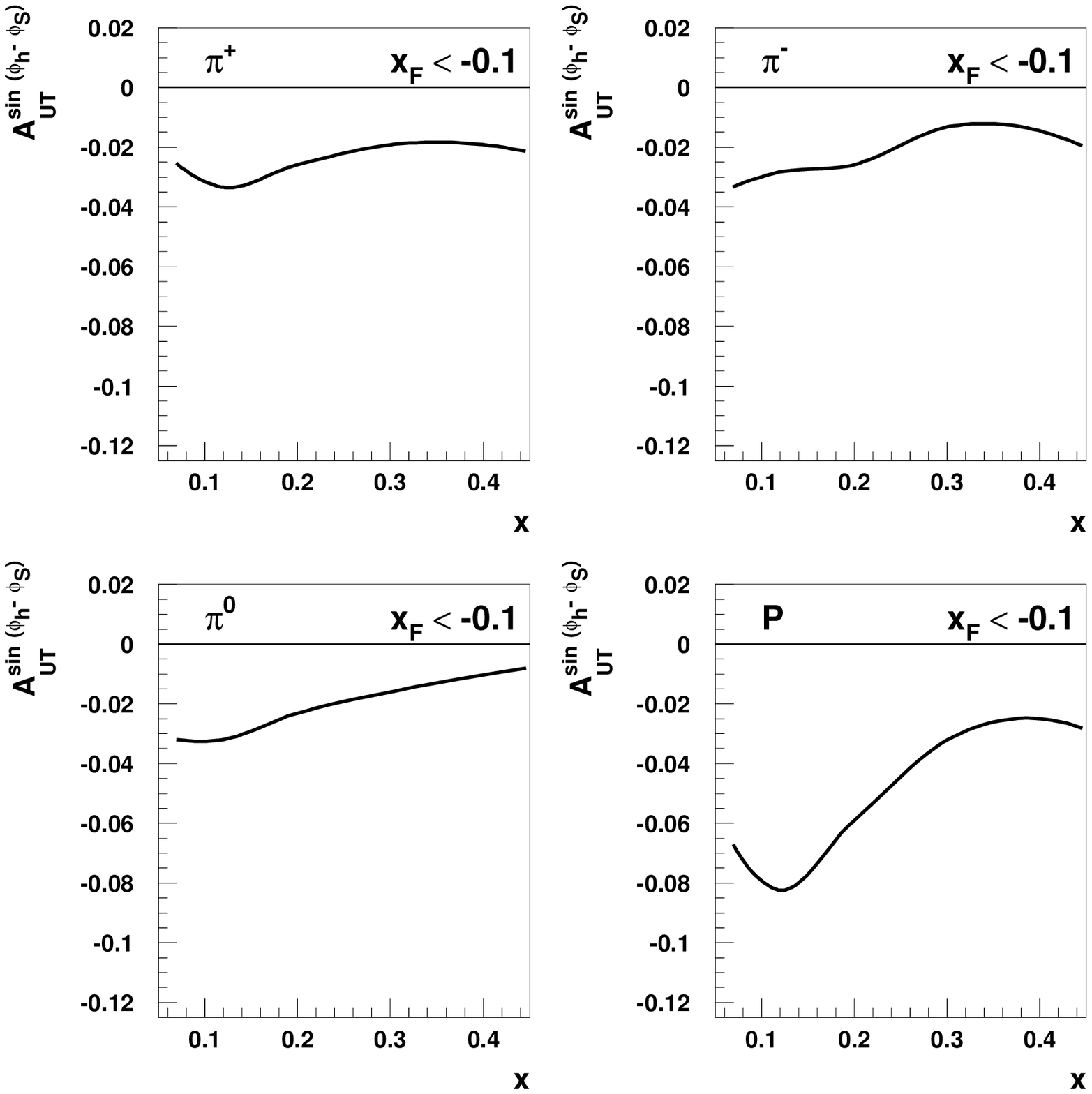}
\vspace {-0.7cm} \caption{\label{fig:sivxb} {\small Predicted
dependence
    of $A_{UT}^{\sin(\varphi_h-\varphi_S)}$ on $x$ for different
    hadrons produced in the TFR ($x_F<-0.1$) of SIDIS of 12 GeV electrons off
    a transversely polarized proton target.}}
\end{center}
\vspace {-0.5cm}
\end{figure}

\begin{figure}[h!]
\begin{center}
\vspace {-0.5cm}
 \includegraphics[width=0.75\linewidth, height=0.65\linewidth]{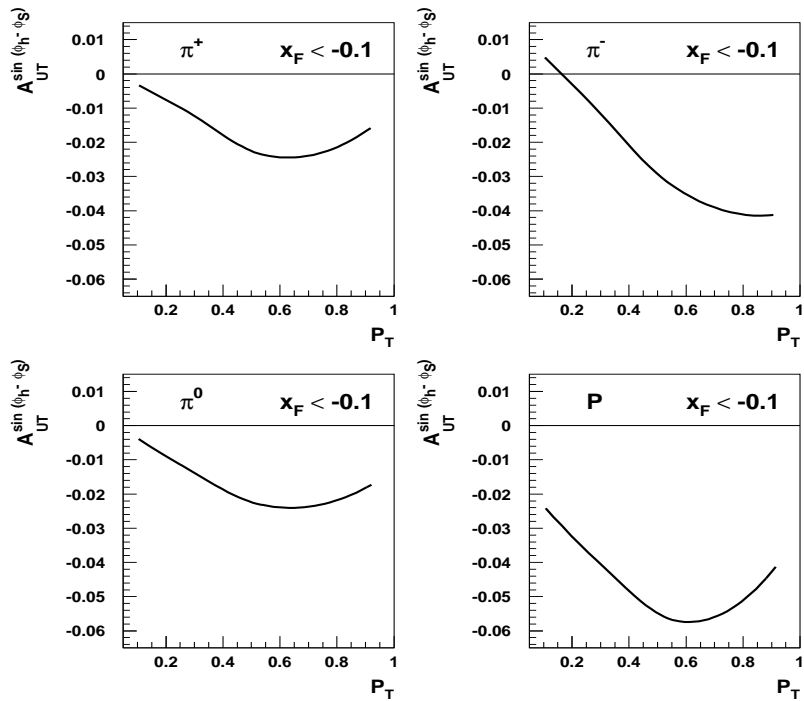}
\vspace {-0.7cm} \caption{\label{fig:sivpt} {\small Predicted
dependence
    of $A_{UT}^{\sin(\varphi_h-\varphi_S)}$ on $p_T$ for different
    hadrons produced in the TFR ($x_F<-0.1$) of SIDIS of 12 GeV electrons off
    a transversely polarized proton target.}}
\end{center}
\vspace {-0.5cm}
\end{figure}

Future facilities as Electron Ion Colliders or upgraded JLab will
have larger kinematic coverage and will offer the possibility of
studying the Sivers effect also with hadrons produced in the TFR.
As an example, the simulations have been done for 12 GeV electron
SIDIS of a proton target. The DIS cut $Q^2>1 (Gev/c)^2$ and $W^2>4
Gev^2$ and a cut on the produced hadron transverse momentum
$P_T>0.05$ GeV/c was imposed. The predictions for $x_F$, $x$ and
$P_T$ dependencies for JLab kinematics are presented on the
Fig.~\ref{fig:sivxf}, Fig.~\ref{fig:sivxb} and
Fig.~\ref{fig:sivpt}, respsctively.

\section{Discussion and Conclusions\label{sec:concl}}

In this article the way of modifying the standard {\tt LEPTO}
event generator in order to include the azimuthal asymmetries
arising from Cahn and Sivers effects is described. Only LO effects
have been taken into account. The azimuthal modulations for Cahn
and Sivers effects have different origins. In the case of Cahn
effect the initial quark transverse momentum is independent of
azimuthal angle but the hard scattering cross section in a non
planar kinematics depends on the final quark azimuthal angle. In
the case of Sivers effect already the initial quark transverse
momentum has an azimuthal modulation. The azimuthal asymmetries
are introduced in both cases by changing the struck quark/string
azimuthal distribution during event generation. The hadronization
part of program ({\tt JETSET}) is left unchanged. The possible
influence of the higher twist distribution functions as well as
possible modifications of hadronization in the case of polarized
target~\cite{ak} have been ignored.

The advantage of this MC based approach compared to standard QCD
factorized approach is in the full coverage of produced hadron
phase space. The modified generator will be useful for complete MC
simulations of experiments including Cahn and Sivers effects both
in the CFR and in the TFR and also for global analysis of these
effects.

Figs.~\ref{fig:cahnemc} and~\ref{fig:sivherm} demonstrate that the
modified {\tt LEPTO} event generator well describing the data in
the CFR both for Cahn and Sivers asymmetries. The description of
Cahn effect in the TFR looks unsatisfactory. This discrepancy can
be explained either by some unaccounted contributions in the TFR
or by insufficient precision of experimental data. One can notice
from the experimental points in Fig.~\ref{fig:cahnemc} that the
integrated value of $\langle \cos \phi_h \rangle$ for charged
hadrons in the CFR is not compensated by that in TFR. It seems
improbable that this imbalance can be compensated by larger values
of $\langle \cos \phi_h \rangle$ of neutral hadrons at $x_F \simeq
-1$.

The new high statistic measurements will allow to check the
predictions of the approach presented here and better understand
the effects of the quark intrinsic transverse momentum and
hadronization mechanism in SIDIS.

\section*{Acknowledgements}

The author express his gratitude to M.~Anselmino, A.~Prokudin for
discussions and also to NUCLEOFIT group members of the General
Physics Department `A.~Avogadro' of the Turin University for
interest in this work.

\end{document}